\documentclass[a4paper]{jpconf}
\usepackage{graphicx}
\begin{document}
\title{Working Group Report on TeV Particle Astrophysics}

\author{Ivone F. M. Albuquerque$^1$, Sergio Palomares-Ruiz$^{2,3}$,
Tom Weiler$^2$}
\address{$^1$Instituto de F\'{i}sica, Universidade de S\~{a}o Paulo, Brazil}
\address{$^2$Department fo Physics \& Astronomy, Vanderbilt University,
   Nashville TN 37235 USA}
\address{$^3$Institute for Particle Physics Phenomenology (IPPP),
  University of Durham, Durham, DH1 3LE, UK}
\ead{ifreire@if.usp.br; sergio.palomares-ruiz@durham.ac.uk;
tom.weiler@Vanderbilt.Edu}

\begin{abstract}
This working group focused mainly on the complementarity among particle physics
and astrophysics. The analysis of data from both fields will better
constrain theoretical models. Much of the discussion focused on detecting
dark matter and susy particles, and on the potential of neutrino 
and gamma-ray astrophysics for seeking or constraining new physics.
\end{abstract}

\section{Introduction}

Complementarity among particle physics and astrophysics was the tone of the
discussions in this working group. Both theoretical and experimental aspects
of this complementarity were discussed. There were eleven talks, three on 
detecting 
supersymmetric particles (more specifically Next to Lightest Supersymmetric
Particles) in km$^3$ neutrino telescopes, a talk on possibilities of finding
new physics with ultra high energy neutrinos, two talks on a mechanism for generating TeV gamma-rays and neutrinos 
(photo-dissociation followed by nuclear de-excitation) largely ignored by the gamma-ray community, 
a talk on better constraining supersymmetric 
parameters when combining astrophysics with LHC data, two talks on
dark matter, a talk on constraining spacetime foam with neutrinos, 
and a talk on a novel way to determine the Higgs field.

\section{Detecting NLSPs with neutrino telescopes}

\indent 
{\bf Zackaria Chacko} reported on his work \cite{abc}, in which  
the possibility of detecting the Next to Lightest Supersymmetric Particle
(NLSP) with neutrino telescopes was first suggested. 
Within susy scenarios in which the breaking scale is on
the lighter side (below $10^{10}$ GeV), the lightest supersymmetric particle
(LSP) is the gravitino. In most models within these scenarios the NLSP
is a charged long lived slepton (typically the right handed $\tilde\tau$). The work reported here 
determines that NLSPs produced in high energy neutrino interactions inside
the Earth can be detected in neutrino telescopes. 

Although the NLSP production 
cross section is low (about three orders of magnitude lower than lepton 
production from the same interactions), the energy loss of these particles
going through the Earth is also very small. This makes the NLSP range large,
typically thousands of kilometers. The large range compensates for the low
production cross section, making possible the NLSP detection.
Moreover, the NLSPs will always be produced in pairs and go through the detector. As its
energy loss is small it will look like a pair of low energy muons going through
the detector. The main background will be dimuon events originating from decay of charmed
particles, themselves produced from neutrino interactions in the Earth. Although
the dimuon background rate is slightly higher than the NLSP's, it can be 
effectively reduced by three cuts: track separation in the detector, a high energy cut,
and a full track requirement \cite{abc2}. 
The conclusion is that km$^3$ neutrino telescopes can detect NLSPs if the
gravitino is the LSP. In this way it will probe the supersymmetry breaking scale
in a fashion complementary to susy measurements at the LHC.

{\bf Markus Ahlers} described his work on the same subject \cite{ahlers} and concluded
that single $\tilde\tau$ events (occuring when one $\tilde\tau$ misses the detector) play
a subdominant role when compared to muon events and will not affect the determination
of the cosmic neutrino flux. He also discussed the possibility of stau trapping in
the detector.

{\bf Mary Hall Reno} presented her work on the $\tilde\tau$ energy loss. The 
radiative energy loss \cite{reno1} was determined.  It is slightly dependent on the
energy. The main contribution is the loss due to photonuclear interactions. A comparison
between the energy loss determined by the scaling of the lepton to the slepton mass
to a direct calculation of each radiative process (pair production, bremsstrahlung and photonuclear)
shows that the scaling calculation overestimates the energy loss by a factor of two.
She also presented work on $\tilde\tau$ weak interactions \cite{reno2}.
The $\tilde\tau$ charge current interaction will remove the particle while the neutral
current contributes to its energy loss. She showed that the neutral current weak
interaction will always be subdominant when compared to the photonuclear energy
loss. The charged current interaction may however be significant for $\tilde\tau$'s
energies above $10^9$ GeV, depending on the particular parameters of the model.

\section{Ultra High Energy Neutrinos as a Window to New Physics}

{\bf Ina Sarcevic} pointed out the importance of ultra high energy neutrinos in
revealing new physics. Since neutrinos are stable and neutral they
point back to their sources without losing energy.  This makes neutrino astronomy
a unique window in looking deep into sources as well as looking to processes which
occured back in the past.

Probable sources of ultra high energy neutrinos are the same as the ones which
produce ultra high energy cosmic rays (``cosmogenic'' or ``GZK'' sources),
Active Galactic Nuclei and Gamma Ray Bursts. Some more peculative sources 
such as topological defects and Z-bursts (at the level of the AGASA rate) 
seem to be incompatible with some data.
One has to bare in mind that new physics might modify neutrinos interactions,
and alter the SM conclusions.
Also, there is a large uncertainty on the ultra high energy neutrino cross section.
Neutrino signatures in various detectors such as neutrino or cosmic
ray telescopes were discussed.

Specific examples of new physics to be probed by neutrinos were discussed.
These were microscopic black holes as predicted by TeV scale gravity models; exchange of Kaluza-Klein
gravitons; production of charged sleptons in neutrino interactions; and electroweak
instanton-induced processes.

\section{TeV Gamma Rays and Neutrinos from Nuclear Photo-dissociation/De-excitation}

It is well known that TeV gamma-rays can originate in two different ways 
in astrophysical sources, either in electromagnetic (electron 
bremsstrahlung, inverse Compton scattering, synchroton radiation) or 
hadronic processes (through $\pi^0$ production in $pp$ or 
$p\gamma$ interactions). The talk by 
{\bf Sergio Palomares-Ruiz} was devoted to a third mechanism: the 
photo-disintegration of nuclei at the source, followed by the 
de-excitation of the daughter nuclei. 
Although discussed some time ago~\cite{BKM87}, this mechanism has been largely 
ignored by the gamma-ray community.

The basic idea is as follows. A highly relativistic nucleus propagates 
in a photon background. If the boost factor of this nucleus is such that 
photo-excitation via the Giant Dipole Resonance (GDR) is possible 
($\epsilon_\gamma \sim 10~{\rm  MeV} - 30~{\rm MeV}$ in the nucleus rest frame), 
then the emission of MeV gamma-rays from the excited daughter nucleus 
(in the rest frame of the nucleus) results. 
To get TeV gamma-rays in the lab from MeV GDR energies in the nuclear frame, 
the boost factor should be $\sim 10^6-10^7$. 
But this implies that in order to reach the GDR energy, the ambient photons must have energies in the far 
ultraviolet; such is expected from Lyman $\alpha$ emissions of hot 
stars. It was also shown that (i) this process effectively sets a lower 
limit on the resulting gamma-ray energy, with no counterpart at lower 
energies, (ii) that a low density interstellar medium is needed in order 
for this process to dominate over the $\pi$ production and decay mode, 
and that (iii) concomitant stripped neutrons will $\beta$-decay to give rise to an 
associated flux of TeV antineutrinos (smaller than the flux of gamma-rays).

As a follow up of Palomares-Ruiz's talk on the nuclei photo-disintegration/de-excitation  
mechanism for TeV gamma-ray production, {\bf Haim Goldberg} applied 
the mechanism to a particular TeV gamma-ray data set for which there is 
no compelling explanation. This is the HEGRA data~\cite{HEGRA} from the 
CygnusOB2 region.  CygnusOB2 is a massive starburst association. 
This region is very rich in hot, young stars, and could 
provide the right ambient photon background for the mechanism sketched 
above. Goldberg showed three important features which arise from the 
characteristics of this region: (i) the dissociation time is larger than 
the diffusion time so that a single-dissociation calculation is valid, 
(ii) the observed gamma-ray flux is expected to have roughly the same 
power index as the source's nuclear flux, 
and (iii) there is a sharp cutoff 
predicted at $\sim$ 0.5 TeV, with no lower energy counterpart.

Goldberg concluded that the required nuclear density must arise from 
high efficiency acceleration of abundant nuclei trapped at 
much lower energies, and that all the energy requirements are satisfied 
by the CygnusOB2 region. Finally, he showed that the calculated CygnusOB2 
neutrino signal from the concomitant neutron decay 
is not statistically significant in the forthcoming IceCube experiment.

\section{Constraining Supersymmetry Parameters with Astro- Plus LHC Data}

{\bf Dan Hooper} reported on the complementarity of astrophysics data and the LHC.
The LHC is aimed at discovering new physics including  
supersymmetric particles. Under the assumption that supersymmetry will stabilize the 
standard model in the TeV region, the LHC will be able to determine the LSP mass within 10\%. 
Depending on the mass scale, it will be possible to determine sleptons,
squark and heavier neutralino masses. However it will be hard to determine various 
supersymmetry parameters, including the bino/wino/higgsino composition of neutralinos. 

Hooper showed that it is possible to determine the neutralino composition
when combining astrophysical measurements with LHC data. 
This is possible because the neutralino's 
annihilation and elastic scattering cross sections are very sensitive to its composition.
Dark matter direct detection will probe the correlation between 
the elastic scattering cross section, the neutralino composition,
and the susy parameters $\tan\beta$, $a$ and \cite{dan1}.
If the elastic scattering is spin dependent, the
constraints from direct detection will remain very weak, and indirect detection
via annihilation might contribute to the determination of the above parameters \cite{dan2}.

As gammas compose the neutralino annihilation products, they might also
bring information on neutralino properties. Gamma ray observations might
contrtibute to determine the fraction of various annihilation products, but
will not help on determining the total cross section. Positron observations
might contribute to pin down the total cross section. The positron flux 
generated in neutralino annihilation is dominated by our Galactic dark matter
distribution.

Hooper also showed that the combination of these results with LHC measurements can not only 
help determine the composition of the neutralino, but can also break the 
degeneracies of the coannihilation/funnel/bulk parameter space, 
and can determine ${\rm m_A}$ for models in the A-funnel region of the parameter space. 

\section{Dark Matter}

\subsection{Bounds on Dark Matter Annihilation Cross Section}

{\bf John Beacom} reported on bounding the dark matter (DM)
annihilation cross section using neutrino fluxes~\cite{john}. 
The size of the annihilation cross-section depends on the DM model. 
For example, if the DM is a thermal relic it will have a very different annihilation 
cross section than if it only appeared or grew mass in the late universe.

Previous work bounded the annihilation cross section from the visible 
annihilation products, especially the resulting diffuse photon flux. 
Beacom reported on an opposite approach, where the bound is 
determined from the ``invisible'' neutrino annihilation products. 
In assuming that all DM annihilates 100\% into
neutrinos, the bound which results closes a loophole, namely the possibility that 
DM annihilation into visible particles is negligible.


As DM traces density square, the determination of the neutrino flux requires
an integration over the DM halo radial density profile, the distribution of
halo masses and their evolution with redshift. 
this is dealt with by adapting 
the prior calculation of the photon bound~\cite{johngam}.
The resulting neutrino signal has a sharp feature, which can be compared 
to the atmospheric neutrino background published by 
AMANDA, Frejus and Super-Kamiokande. This comparison allows the 
determination of an upper bound on the DM annihilation cross section.
By closing the loophole mentioned above, 
the bound rules out the KKT \cite{kkt} model of structure formation,
which proposed DM self annihilation to soften halo cusps.
Although the new bound does not reach the natural scale of thermal relics,
it can be improved as new data appears from Super-Kamiokande and IceCube.

\subsection{Squishing Dark Matter with Black Holes}

{\bf Gianfranco Bertone} reported on enhancement of dark matter (DM) annihilation
due to Black Holes. The idea is that DM mini-halos will react to
the formation or adiabatic growth of Intermediate Black Holes (IBH) annihilating
into mini-spikes. He showed that the adiabatic growth of a massive object at
the center of a DM distribution will induce redistribution of matter.
If the DM density profile follows a power law with spectral index $\gamma$,
the DM will be redistributed into a new power law with index 
$(9 - 2\gamma)/(4 - \gamma)$.  Near $r=0$ a DM mini-spike will occur.

DM annihilation at the mini-spikes will produce bright gamma-rays \cite{gf1}. 
Each IBH spike can be as bright as the whole galaxy.
This radiation can be detected by Air Cherenkov Telescopes
such as Cangaroo, Hess, Magic and Veritas, if they extend their observations
to higher energies.  
Better is a full sky survey, such as will happen with GLAST.
However, GLAST is not sensitive to energies
above 300 GeV and so might miss the DM detection if the DM mass is heavy.
Bertone showed the expected flux as a function of the number of black holes for
Glast and EGRET \cite {gf1}.

Another possibility for finding DM mini-spikes is with neutrino telescopes.
It was argued that mini-spikes can also be bright sources of neutrinos.
The rates expected for ANTARES and IceCube are encouraging~\cite{gf2},
with ANTARES being better located since it can see the Galactic center. 

\section{Spacetime Foam}

{\bf Luis Anchordoqui} reported on a study using the flavor ratios of incoming 
cosmic neutrinos at IceCube to improve bounds on the scale of
quantum gravity, $M_{\rm QG}$.
It is often stated that the quantum effects of gravity may never be
experimentally accessible because they would be manifest only at the
Planck scale, $M_{\rm Pl} \equiv \sqrt{\frac{\hbar c}{G_{\rm N}}}
\simeq 1.2 \times10^{19}$~GeV. However, being a
non-renormalizable interaction, gravity 
may leave a distinctive imprint at energies much lower than
the Planck scale, akin to observable parity violation in
$\beta$-decay at energies far below the $G_F^{-1/2}$ scale of the 
weak interaction. For example, if
quantum space-time has a `foamy' structure in which Planck length size
black holes form and evaporate on the Planck time scale, then there
may be a loss of quantum information across their event horizons 
which induces decoherence of a pure state.
The particle most sensitive to such effects would appear to be the
pure neutrino state which gives rise to neutrino flavor oscillations.

A heuristic view of decoherence induced by neutrino interactions with 
the virtual black holes is that flavor
is randomized by these interactions, since black holes
are believed not to conserve global quantum numbers.
The visible effect of quantum decoherence, then, would be to alter the
neutrino flavor ratios $\nu_e:\nu_\mu:\nu_\tau$ to
$\simeq 1:1:1,$ regardless of the initial flavor content.
Since the decoherence effects grow with the distance traveled by the (anti) 
neutrinos,
observation of a cosmic 
$\nu$-flux with flavor ratios $\neq 
1:1:1$ could place strong constraints on the energy
scale of quantum decoherence.
It has recently been suggested~\cite{Hooper:2004xr} that antineutrinos
originating in the decay of neutrons from photo-dissociated cosmic ray sources
in the Galaxy~\cite{Anchordoqui:2003vc} can provide a very sensitive 
probe of foam dynamics.
Present limits are $M_{\rm QG} > 10^{16}$~GeV from $pp$-annihilation 
into a BH,
and $M_{\rm QG} > 10^{28/(n-1)}$~GeV from SuperKamiokande flavor 
measurements, where $n$ is the power of the decoherence parameterization
$\sim M_{\rm QG}\,(E/M_{\rm QG})^n$~\cite{Anchordoqui:2005is}.
Anchordoqui claimed that IceCube will dramatically 
improve the sensitivity to decoherence effects~\cite{anchorQG}.
With future data, perhaps from Galactic neutron$\rightarrow {\bar \nu}_e$ 
sources, IceCube can reach $M_{\rm QG} > 10^{46/(n-1)}$~GeV.

\section{Higgs Field}

{\bf Stephen Reucroft} proposed a novel way~\cite{higgs} to detect the Higgs sector. His idea is
to look for the Higgs field instead of looking for the Higgs particle.
His argument is that at short range, such as exists in momentarily is pair production,
the Higgs Yukawa potential alters the effective masses of the pair particles.
He proposes to look for an effect in the threshold cross section of massive pair
production.  The reaction $e^+ e^-\rightarrow W^+ W^-$ provides an optimized laboratory.

In the group discussion that followed his report, it was argued that this effect is small,
and debated whether radiative corrections already include this effect.

\end{document}